\def\rfr#1{eq. (\ref{#1})}

\def\derp#1#2{\rp{\partial{#1}}{\partial{#2}}}

\def\virg#1{``#1''}
\def\cf#1#2{\dot\Omega^{\rm #2}_{.#1}}

\def\eqi{\begin{equation}}
\def\eqf{\end{equation}}
\def\eqia{\begin{eqnarray}}
\def\eqfa{\end{eqnarray}}
\def\rp#1#2{{#1\over#2}} \def\lb#1{\label{#1}}

\def\bds#1{\boldsymbol{#1}}

\documentclass{aastex}
\usepackage{url}
\usepackage{amsmath,amsthm,amscd,amssymb}
\usepackage{latexsym}
\usepackage{graphicx,epsfig}

\RequirePackage{color}

\newcommand{\emaila}{lorenzo.iorio@libero.it}

\begin{document}

\title{How accurate is the cancelation of the first even zonal harmonic of the geopotential in the present and future LAGEOS-based Lense-Thirring tests?}
\shortauthors{L. Iorio}

\author{Lorenzo Iorio\altaffilmark{1} }
\affil{Ministero dell'Istruzione, dell'Universit\`{a} e della Ricerca (M.I.U.R.). Fellow of the Royal Astronomical Society (F.R.A.S.): Viale Unit\`{a} di Italia 68, 70125, Bari (BA), Italy.}

\email{\emaila}

\begin{abstract}
The strategy followed so far in the performed or proposed tests of the general relativistic Lense-Thirring effect in the gravitational field of the Earth with
laser-ranged satellites of LAGEOS type relies upon the cancelation of the disturbing huge precessions induced by the first even zonal harmonic coefficient $J_2$ of the multipolar expansion of the Newtonian part of the terrestrial gravitational potential by means of suitably designed linear combinations of the nodes $\Omega$ of more than one spacecraft. Actually, such a removal does depend on the accuracy with which the coefficients of the combinations adopted can be realistically known. Uncertainties of the order of 2 cm in the semimajor axes $a$ and 0.5 milliarcseconds in the inclinations $I$ of LAGEOS and LAGEOS II, entering the expression of the coefficient $c_1$ of the combination of their nodes used so far, yield  an uncertainty $\delta c_1=1.30\times 10^{-8}$. It gives an imperfectly canceled $J_2$ signal of 10.8 milliarcseconds per year corresponding to $23\%$ of the Lense-Thirring signature. Uncertainties of the order of $10-30$ microarcseconds in the inclinations yield $\delta c_1=7.9\times 10^{-9}$ which corresponds to an uncanceled $J_2$ signature of $6.5$ milliarcseconds per year, i.e. $14\%$ of the Lense-Thirring signal. Concerning a future LAGEOS-LAGEOS II-LARES combination with coefficients $k_1$ and $k_2$, the same uncertainties in $a$ and the less accurate uncertainties in $I$ as before yield $\delta k_1=1.1\times 10^{-8}, \delta k_2=2\times 10^{-9}$; they imply a residual $J_2$ combined precession of $14.7$ milliarcseconds per year corresponding to $29\%$ of the Lense-Thirring trend. Uncertainties in the inclinations at  $\approx 10$ microarcseconds level give $\delta k_1=5\times 10^{-9}, \delta k_2=2\times 10^{-9}$; the uncanceled $J_2$ effect is 7.9 milliarcseconds per year, i.e. $16\%$ of the relativistic effect.
 \end{abstract}


\section{Introduction}
According to the Einsteinian General Theory of Relativity (GTR), the Lense-Thirring\footnote{According to \citet{Pfi}, it would be more correct to speak about an Einstein-Thirring-Lense effect. } precession of the longitude of the ascending node\footnote{It is an angle in the $\{xy\}$ reference plane, coinciding with the equatorial plane of the central body, between the reference $x$ axis and the line of the nodes, which is the intersection between the test particle's orbital plane and the reference plane.} $\Omega$ of a test particle orbiting a central, slowly rotating body of mass $M$ and angular momentum $S$ is \citep{LT}
\eqi\dot\Omega_{\rm LT}=\rp{2GS}{c^2 a^3(1-e^2)^{3/2}},\lb{ltnodo}\eqf
where $G$ is the Newtonian constant of gravitation, $c$ is the speed of light in vacuum, $a,e$ are the semimajor axis and the eccentricity, respectively, of the test particle's orbit; note that \rfr{ltnodo} does not depend on the inclination $I$ of the orbit to the central body's equator.
 The Lense-Thirring effect is a consequence of the fact that, in its weak-field and slow-motion approximation, the Einstein's field equations of GTR get linearized, thus resembling the linear equations of the Maxwellian electromagnetism. In such a framework, analogously to the magnetic field induced by electric currents acting on a moving electric charge through the Lorentz force, mass-energy currents like those of an isolated rotating mass generate a gravitomagnetic field in the space surrounding it which acts on a moving test particle with a non-central, Lorentz-like force perturbing its Keplerian motion.

Attempts to detect the Lense-Thirring node precessions in the gravitational field of the Earth have been performed so far with the passive geodetic LAGEOS and LAGEOS II satellites \citep{Ciu09}  tracked with the Satellite Laser Ranging (SLR) technique\footnote{It allows to precisely measure the range $\rho$ between a laser station and a spacecraft that is equipped with retroreflectors like just the LAGEOS satellites. The range is deduced from the elapsed time of  flight for
a pulse of laser light traveling from the ground station to the
satellite and back again.} \citep{SLR}; a total accuracy of the order of approximately $10\%$ is claimed \citep{Ciu09}. A further LAGEOS-like SLR target, named LARES, should be launched in late\footnote{See on the WEB: http://spaceflightnow.com/tracking/index.html} 2011 with a VEGA rocket \citep{Ciu09}; its proponent claims that, in conjunction with the existing LAGEOS and LAGEOS II, it will be able to reach a $\approx 1\%$ accuracy in measuring the Lense-Thirring effect \citep{Ciu09}. The values of the relevant orbital parameters and of the Lense-Thirring node precessions for LAGEOS, LAGEOS II and LARES are in Table \ref{OSIGNUR}. They amount to a few ten-hundred milliarcseconds per year (mas yr$^{-1}$ in the following) corresponding to  linear shifts of about $2-5$ m per year at their altitudes.
\begin{table}[t]
\caption{Orbital parameters and Lense-Thirring node precessions of LAGEOS, LAGEOS II and LARES  for
$S_{\oplus} = 5.86\times 10^{33}$ kg m$^2$ s$^{-1}$ \protect\citep{IERS}.  The semimajor axis $a$ is in km, the inclination $I$ is in deg, and the Lense-Thirring rate $\dot\Omega_{\rm LT}$ is in mas yr$^{-1}$.}\label{OSIGNUR}
\begin{tabular}{@{}lllll}
\hline
Satellite & $a$  & $e$ & $I$  & $\dot\Omega_{\rm LT}$  \\
\hline
LAGEOS & 12270 &  0.0045 & 109.9 & 30.7\\
LAGEOS II & 12163 &  0.014 & 52.65 & 31.5 \\
LARES & 7828 &  0.0 & 71.5 & 118.1 \\
\hline
\end{tabular}
\end{table}

Actually, the nodes of such satellites are affected by much larger secular precessions $\dot\Omega_{J_{\ell}}\doteq \dot\Omega_{.\ell}J_{\ell}$ induced by the even ($\ell=2,4,6,...$) zonal ($m=0$) harmonic coefficients $J_{\ell}, \ell=2,4,6,...$ of the multipolar expansion of the Newtonian part of the terrestrial gravitational potential which account for the departures from spherical symmetry of the Earth because of its diurnal rotation \citep{Tap}. The even zonal harmonics, defined as $J_{\ell}\doteq -\sqrt{2\ell +1}\ \overline{C}_{\ell 0}, \ell=2,4,6,...$ in terms of the normalized Stokes coefficients $\overline{C}_{\ell 0}, \ell=2,4,6,...$ \citep{Tap}, are directly estimated as solve-for parameters in global Earth's gravity field solutions\footnote{They are publicly available on the WEB at http://icgem.gfz-potsdam.de/ICGEM/.} obtained by processing huge data sets from dedicated satellite-based mission like CHAMP\footnote{See on the WEB: http://www-app2.gfz-potsdam.de/pb1/op/champ/} and, especially, GRACE\footnote{See on the WEB: http://www-app2.gfz-potsdam.de/pb1/op/grace/index$\_$GRACE.html}.
The most effective even zonals in perturbing the satellites' nodes are the low-degree ones; the coefficients $\dot\Omega_{.\ell}$ of the node precessions for $\ell=2,4$
are \citep{Ciu96,Ior03}
\begin{equation}
\begin{array}{lll}
   \dot\Omega_{.2}= -\rp{3}{2}n\left(\rp{R}{a}\right)^2\rp{\cos I}{(1-e^2)^2}, \\  \\
  \dot\Omega_{.4} = \dot\Omega_{.2}\left[\rp{5}{8}\left(\rp{R}{a}\right)^2\rp{1+\rp{3}{2}e^2}{(1-e^2)^2}\left(7\sin^2 I-4\right)\right] \lb{cf4},
\end{array}
\end{equation}
where $n\doteq\sqrt{GM/a^3}$ is the satellite's Keplerian mean motion and $R$ is the mean equatorial radius of the central body; contrary to the Lense-Thirring precession of \rfr{ltnodo}, the classical precessions of \rfr{cf4} depend on the inclination $I$.
The node precessions due to the first even zonal $J_2$ for LAGEOS, LAGEOS II and LARES are listed in Table \ref{OSIGNUR2}.
\begin{table}[t]
\caption{Node precessions $\dot\Omega_{J_2}\doteq\dot\Omega_{.2} J_2$, in mas yr$^{-1}$, of LAGEOS, LAGEOS II and LARES due to $J_2$. We used $\overline{C}_{20}=-4.841692151273\times 10^{-4}$ from the ITG-Grace2010s  \citep{ITG} global solution. Recall that $J_{\ell}\doteq -\sqrt{2\ell +1}\ \overline{C}_{\ell 0}, \ell=2,4,6,...$.}\label{OSIGNUR2}
\begin{tabular}{@{}llll}
\hline
 & LAGEOS  & LAGEOS II & LARES    \\
\hline
$\dot\Omega_{J_2}$ (mas yr$^{-1}$) & $4.516313623\times 10^8$ & $-8.303250890\times 10^8 $ & $-2.0298203351\times 10^9$\\
\hline
\end{tabular}
\end{table}
It can be noted that they are 7 orders of magnitude larger than the Lense-Thirring precessions of Table \ref{OSIGNUR}.

Thus, suitable linear combinations of the nodes of more than one satellite have been set up in order to purposely cancel out, by construction, the impact of one or more even zonals according to a strategy put forth by \citet{Ciu96}. In particular, the tests performed so far have been conducted with the following LAGEOS-LAGEOS II combination\footnote{See also \citet{Ries,Pavlis}.} \citep{IorioMGM}
\begin{equation} \textcolor{black}{f^{(\rm 2L)}\doteq}\dot\Omega^{\rm LAGEOS}+
c_1\dot\Omega^{\rm LAGEOS\ II }, \lb{combi}\end{equation}
where \begin{equation} c_1\doteq-\rp{\dot\Omega^{\rm LAGEOS}_{.2}}{\dot\Omega^{\rm
LAGEOS\ II }_{.2}}=-\rp{\cos I_{\rm LAGEOS}}{\cos I_{\rm LAGEOS\
II}}\left(\rp{1-e^2_{\rm LAGEOS\ II}}{1-e^2_{\rm
LAGEOS}}\right)^2\left(\rp{a_{\rm LAGEOS\ II}}{a_{\rm LAGEOS}}\right)^{7/2}.\lb{coff}\end{equation}
The future combination involving LARES as well, designed to remove the effect of $J_2$ and $J_4$, is \citep{IorioNA}
\eqi \textcolor{black}{f^{(\rm 3L)}\doteq}\dot\Omega^{\rm LAGEOS}+
k_1\dot\Omega^{\rm LAGEOS\ II }+k_2\dot\Omega^{\rm LARES },\lb{combilares}\eqf
in which
\begin{equation}
\begin{array}{lll}
k_1 = \rp{\cf 2{LARES}\cf4{LAGEOS}-\cf 2{LAGEOS}\cf 4{LARES}}{\cf 2{LAGEOS\ II}\cf 4{LARES}-\cf 2{LARES}\cf 4{LAGEOS\ II}},\\\\
k_2 =  \rp{\cf 2{LAGEOS}\cf4{LAGEOS\ II}-\cf 2{LAGEOS\ II}\cf 4{LAGEOS}}{\cf 2{LAGEOS\ II}\cf 4{LARES}-\cf 2{LARES}\cf 4{LAGEOS\ II}}.
\end{array}\lb{cofis}
 \end{equation}
 It is analogous to the combination of the nodes of LAGEOS and LAGEOS II and the perigee $\omega$ of LAGEOS II \citep{Ciu96} used in the earlier tests \citep{Ciu98}: the coefficients of the precessions of the perigee of LAGEOS II have to be replaced by those of the precessions of the node of LARES.
Table \ref{OSIGNUR}, \rfr{cf4}, \rfr{coff}, and  \rfr{cofis} yield the numerical values of $c_1, k_1,k_2$ shown in Table \ref{OSIGNUR3}.
\begin{table}[t]
\caption{Nominal values of the coefficients $c_1$ of the present LAGEOS-LAGEOS II combination, and $k_1$, $k_2$ of the future LAGEOS-LAGEOS II-LARES combination according to  Table \ref{OSIGNUR}. The combined Lense-Thirring node precessions are $47.8$ mas yr$^{-1}$ (LAGEOS-LAGEOS II, \rfr{combi}), and $50.8$ mas yr$^{-1}$ (LAGEOS-LAGEOS II-LARES, \rfr{combilares}), respectively.}\label{OSIGNUR3}
\begin{tabular}{@{}lll}
\hline
$c_1$ & $k_1$ & $k_2$ \\
\hline
$0.5439211320$ & $0.3603291106$ & $0.0751007658$\\
\hline
\end{tabular}
\end{table}
Table \ref{OSIGNUR3}, \rfr{combi} and \rfr{combilares} yield 47.8 mas yr$^{-1}$ and 50.8 mas yr$^{-1}$, respectively, for the predicted Lense-Thirring combined precessions.
The LAGEOS-LAGEOS II combination of \rfr{combi} is fully affected by the node precessions of degree higher than 2, i.e. $\ell=4,6,8,...$; instead, the LAGEOS-LAGEOS II-LARES combination of \rfr{combilares} will be fully impacted by the even zonals of degree higher than 4, i.e. $\ell=6,8,...$. A realistic evaluation of the systematic uncertainty induced by the mismodeling in such uncanceled even zonals on the predicted Lense-Thirring signals\textcolor{black}{, i.e. \eqi \left.\delta f^{(q\rm L)}\right|_{J_{\ell}}\leq\sum_{\ell=2}\left|\rp{\partial{f^{(q\rm L)}}}{\partial J_{\ell}}\right|\delta J_{\ell}, q=2,3,\eqf } has been the subject of several recent studies summarized in \citet{Iorio09}. Concerning the present-day LAGEOS-LAGEOS II tests \citep{Iorio09}, the total accuracy may be up to $2-3$ times larger than claimed by \citet{Ciu09}; in the case of the future tests involving LARES, both gravitational \citep{Iorio09} and non-gravitational \citep{IorioAPPB} mismodeled perturbations should likely impact the mission at a level larger than the claimed $1\%$.

In this paper we want to deal with another, subtle issue pertaining the systematic bias induced by the even zonal harmonics of the geopotential\footnote{I thank an anonymous referee of a previous paper of mine for having pointed out this issue to me.}. Indeed, all the studies performed so far relied upon the assumption of a perfect cancelation of $J_2$ by the combinations of \rfr{combi} and \rfr{combilares}. Actually, it depends on the accuracy with which their coefficients $c_1, k_1,k_2$ can be known; given the huge magnitude of the nominal $J_2$-induced precessions of Table \ref{OSIGNUR2} with respect to the gravitomagnetic ones of Table \ref{OSIGNUR}, it has to be quite high to really allow for a measurement with a given level of uncertainty $X\%$. Instead, until now, the coefficients of the combinations of \rfr{combi} and \rfr{combilares} have always been computed with a few decimal digits. In other words, one has to evaluate
\eqi \textcolor{black}{\left.\delta f^{(\rm 2L)}\right|_{c_1}\leq}\delta c_1\left|\dot\Omega^{\rm LAGEOS\ II}_{J_2}\right|\eqf and
\eqi \textcolor{black}{\left.\delta f^{(\rm 3L)}\right|_{k_1,k_2}\leq}\delta k_1\left|\dot\Omega^{\rm LAGEOS\ II}_{J_2}\right|+\delta k_2\left|\dot\Omega^{\rm LARES}_{J_2}\right|\eqf
as further sources of systematic uncertainty with respect to the combined Lense-Thirring precessions \textcolor{black}{which have to be added to $\left.\delta f^{(q\rm L)}\right|_{J_{\ell}},\ q=2,3$. In order to avoid possible confusions and misunderstandings, it should be clarified that it would be incorrect to evaluate the impact of the uncertainties in the combinations' coefficients by only taking terms proportional  to  cross products of the errors like $\delta c_1\delta J_{\ell},\delta k_1\delta J_{\ell},\delta k_2\delta J_{\ell}$ instead of those proportional to $\delta c_1,\delta k_1,\delta k_2$ themselves, as done by us. Indeed, it is well known from elementary theory of errors that if an empirically determined quantity $f$ depends on several parameters $p_j, j=1,2...$ affected by uncertainties $\delta p_j$, the total uncertainty in $f$ is just \eqi \delta f\leq \sum_{j=1}\left|\rp{\partial f}{\partial p_j}\right|\delta p_j.\eqf Mixed terms of the form \eqi \derp f {p_i}\derp f {p_j}\sigma_{p_j p_j}\eqf appear only in case of a correlation, which is absent in the present case. Indeed, the coefficients $c_1,k_1,k_2$ of the combinations $f^{(q{\rm L})}, q=2,3$ and the even zonals $J_{\ell}$ of the geopotential are not solved-for parameters\footnote{Actually, the Lense-Thirring effect itself has never been explicitly modelled and solved-for in all the analyses performed so far.}, simultaneously estimated in the same global solution: otherwise, one may look at their mutual correlations in the covariance matrix. Anyway, even if it was the case, a conservative evaluation of the total uncertainty would require to neglect the covariance by only retaining the linear sum of the individual mismodelled terms.
}

The paper is organized as follows.
In Section \ref{due} we will deal with the ongoing LAGEOS-LAGEOS II tests. The LAGEOS-LAGEOS II-LARES case will be tackled in Section \ref{tre}, while Section \ref{quattro} contains the summary and the conclusions.
\section{The LAGEOS-LAGEOS II case}\lb{due}
The coefficient $c_1$ of \rfr{coff} actually depends on the semimajor axes, the eccentricities and the inclinations of both LAGEOS and LAGEOS II. Thus, the accuracy with which it is possible to know it is set by the uncertainties in such Keplerian orbital elements. They are not directly measurable quantities being, instead, computed from the satellite's state vectors $\bds r$ and $\bds v$ whose components are, in turn, estimated in a least-square sense by processing the differences between the observed
and calculated ranges at different times \citep{Tap}.

Let us, now, consider in detail how to assess the uncertainty  in the semimajor axis $a$ due to a key geodetic parameter, i.e. the Earth's gravitational parameter $GM$ which must be assumed as known to pass from the state vector to the Keplerian orbital elements.
For a Keplerian orbit the semimajor axis is given by
\eqi a = \left(\rp{2}{r}-\rp{v^2}{GM}\right)^{-1},\lb{sma}\eqf
where $r$ and $v$ are the satellite's geocentric distance and speed, respectively.
Thus, the relative uncertainty in $a$ due to $GM$
is
\eqi \left.\rp{\delta a}{a}\right|_{GM}=\left(\rp{\delta GM}{GM}\right)\rp{v^2}{GM}a=\left(\rp{\delta GM}{GM}\right)\rp{1+e^2+2e\cos f}{1-e^2}.\eqf
Averaging over one orbital period $P_{\rm b}\doteq 2\pi/n$ by means of
\eqi \rp{dt}{P_{\rm b}} = \rp{(1-e^2)^{3/2}}{2\pi(1+e\cos f)^2},\eqf
it turns out that
\eqi \left\langle\left.{\delta a}\right|_{GM}\right\rangle=\left(\rp{\delta GM}{GM}\right) a.\eqf
Since \citep{IERS}
\eqi \rp{\delta GM}{GM}=2.00702\times 10^{-9},\eqf
the average uncertainties in the semimajor axes of LAGEOS, LAGEOS II and LARES are of the order of
\begin{equation}
\begin{array}{lll}
  \left\langle\delta a_{\rm LAGEOS}\right\rangle \leq   2.5\ {\rm cm}, \\  \\
  \left\langle\delta a_{\rm LAGEOS\ II}\right\rangle  \leq 2.4\ {\rm cm}, \\  \\
  \left\langle\delta a_{\rm LARES}\right\rangle  \leq  1.6\ {\rm cm}\lb{da}.
\end{array}
\end{equation}
 An issue is that the Earth's gravitational parameter $GM$ is estimated by processing long SLR data sets in which just LAGEOS and LAGEOS II play a fundamental role \citep{Dunn}. Moreover, the gravitomagnetic field of the Earth has never been accounted for  in the solutions yielding $GM$ produced so far, so that a twofold source of a-priori \virg{imprinting} of the Lense-Thirring itself is present in the values of the Earth's $GM$ adopted. It would be necessary to use figures obtained without including  data from SLR targets, especially LAGEOS and LAGEOS II, although they may be less accurate.

Actually, the total, realistic uncertainty in $a$ should be even larger because of $r$ and $v$ entering \rfr{sma}. Indeed, concerning the uncertainty in $r$, it includes the\footnote{Of course, it has to be intended in the root$-$mean$-$square sense; it is not the mere single$-$shot mm$-$accuracy.} cm$-$level accuracy in the station$-$satellite range $\rho$ and the uncertainty in the geocenter$-$station position $R_{\rm sta}$, of the order of about $1-2$ cm \citep{STA}.
Anyway, in future calculation we will use the values of \rfr{da}.

Concerning the LAGEOS-LAGEOS II combination of \rfr{combi} used for the present-day tests, the uncertainty in its coefficient $c_1$ can be conservatively evaluated as
\eqi \delta c_1\leq \left|\derp{c_1}{a_{\rm L}}\right|\delta a_{\rm L}+\left|\derp{c_1}{a_{\rm L\ II}}\right|\delta a_{\rm L\ II}+\left|\derp{c_1}{I_{\rm L}}\right|\delta I_{\rm L}+\left|\derp{c_1}{I_{\rm L\ II}}\right|\delta I_{\rm L\ II}.\lb{savc1}\eqf
If, together with \rfr{da} for the uncertainties in the semimajor axes, we assume a reasonable and realistic value for the uncertainties in the inclinations of LAGEOS and LAGEOS II, i.e.\footnote{Indeed, it corresponds to a reasonable $\delta r\approx 1+2=3$ cm from $\delta I\approx \delta r/a$.} $\delta I=0.5$ mas, \rfr{savc1} yields
\eqi \delta c_1=1.30\times 10^{-8}\eqf corresponding to a residual $J_2$ bias (see Table \ref{OSIGNUR2})
\eqi \delta c_1 \left|\dot\Omega_{J_2}^{\rm LAGEOS\ II}\right|= 10.8\ {\rm mas\ yr}^{-1}\eqf and, thus to a percent uncertainty in the Lense-Thirring combined signal of $23\%$. Instead, if we consider\footnote{Such figures seem to be unrealistic because they would imply an accuracy $\delta r\approx a\delta I$ in
reconstructing the orbits of LAGEOS and LAGEOS II, on average, of 0.2 cm
and 0.06 cm, respectively.}
$\delta I_{\rm L}=30\ \mu{\rm as}$, $\delta I_{\rm L\ II}=10\ \mu{\rm as}$ claimed by \citep{Ciu09} we have
\eqi \delta c_1=7.9\times 10^{-9}\eqf yielding an uncanceled $J_2$ signal (see Table \ref{OSIGNUR2}) \eqi\delta c_1\left|\dot\Omega_{J_2}^{\rm LAGEOS II}\right|= 6.5\ {\rm mas\ yr}^{-1}\eqf  which corresponds to a percent uncertainty of $14\%$.

These results show that the issue of the imperfect cancelation of the largest node precessions due to $J_2$ cannot be neglected in the evaluation of the total error budget, especially because the previous figures have to be added to those accounting for the mismodeling in the other even zonal harmonics of higher degree fully impacting the combination of \rfr{combi}.

\section{The LAGEOS-LAGEOS II-LARES case}\lb{tre}
The case of the LAGEOS-LAGEOS II-LARES combination of \rfr{combilares} can be treated in a similar way. \citet{IorioAPPB} preliminarily dealt with it by considering the impact of $a_{\rm LR}$ and $I_{\rm LR}$ only on $k_2$. Instead, one has to fully take into account  the uncertainties of the orbital elements of LAGEOS and LAGEOS II as well in both $k_1$ and $k_2$ according to
\eqi \delta k_{1/2}\leq \sum_i\left|\derp{k_{1/2}}{\psi_i}\right|\delta \psi_i,\ \psi_i = a_{\rm L},I_{\rm L},a_{\rm L\ II},I_{\rm L\ II},a_{\rm LR},I_{\rm LR}.\eqf
By using \rfr{da} for $\delta a$ and assuming $\delta I=0.5$ mas for LAGEOS, LAGEOS II and LARES, the uncertainties in $k_1$ and $k_2$ are
\eqi\delta k_1 = 1.1\times 10^{-8},\ \delta k_2 = 2\times 10^{-9},\eqf
which yield an uncanceled $J_2$ signal
\eqi\delta k_1\left|\dot\Omega_{J_2}^{\rm LAGEOS\ II}\right|+\delta k_2\left|\dot\Omega_{J_2}^{\rm LARES}\right| = 14.7\ {\rm mas\ yr}^{-1}.\eqf
It corresponds to $29\%$ of the combination of the Lense-Thirring node precessions.
If, instead, in addition to \rfr{da} one adopts   $\delta I_{\rm L}=30\ \mu{\rm as}$, $\delta I_{\rm L\ II}=10\ \mu{\rm as}$ \citep{Ciu09} and, say, $\delta I_{\rm LR}=20\ \mu{\rm as}$ the uncertainties in $k_1$ and $k_2$ are
\eqi \delta k_1 = 5\times 10^{-9},\ \delta k_2 = 2\times 10^{-9}. \eqf
They yield a residual $J_2$ signature
\eqi\delta k_1\left|\dot\Omega_{J_2}^{\rm LAGEOS\ II}\right|+\delta k_2\left|\dot\Omega_{J_2}^{\rm LARES}\right| = 7.9\ {\rm mas\ yr}^{-1},\eqf
amounting to $16\%$ of the predicted Lense-Thirring trend.

Concerning the imperfectly canceled $J_4$ signal, it turns out that it is of no concern amounting to $0.006-0.008$ mas yr$^{-1}$.

Thus, independently of the lingering uncertainty in how to realistically assess the bias due to the mismodeling in the uncanceled even zonal harmonics of higher degree impacting in full the combination of \rfr{combilares}, the imperfect removal of the effect of $J_2$ alone is sufficient to make dubious the achievement of the goal of a $\approx 1\%$ total accuracy in the future LAGEOS-LAGEOS II-LARES tests.
\section{Summary and conclusions}\lb{quattro}
One of the major sources of systematic uncertainty in the measurement of the gravitomagnetic Lense-Thirring precessions of the nodes $\Omega$ of the laser-tracked LAGEOS-type  satellites in the gravitational field of the Earth is given by the much larger competing classical node precessions induced by the even zonal harmonic coefficients $J_{\ell}, \ell=2,4,6,...$ of the expansion in multipoles of the non-spherically symmetric terrestrial gravitational potential.
The strategy followed so far to partially circumvent such an issue consisted of suitably designing linear combinations of the nodes of more than one satellite to cancel out, by construction, the effects of $J_2$, as in the ongoing LAGEOS-LAGEOS II test, and of $J_4$ as well, as in the future LAGEOS-LAGEOS II-LARES scenario.
In addition to the usual systematic uncertainty due to the mismodeling in the even zonals of higher degree which fully impact such combinations, another source of non-negligible uncertainty of gravitational origin has to be taken into account. It is due to the imperfect cancelation of the effects of  $J_2$ because of the uncertainty in the coefficients entering the combinations set up just to remove it. Indeed,
the numerical values of such coefficients, released with just a few decimal digits so far, explicitly depend on the numerical values  of the semimajor axes $a$, the inclinations $I$ and the eccentricities $e$ of the satellites involved. Thus, the uncertainties with which such Keplerian orbital elements are known unavoidably have repercussions onto the coefficients themselves.
For uncertainties in the semimajor axes of $1-2$ cm and of about $0.5-0.01/0.03$ milliarcseconds in  the inclinations we have shown that the resulting systematic bias due to the imperfect removal of the $J_2$ signal may be as large as   $14-29\%$ of the Lense-Thirring signatures.
\textcolor{black}{\section*{Acknowledgements}
I gratefully thank an anonymous referee for her/his valuable critical remarks which helped in better elucidating a subtle issue.}

\end{document}